\documentclass[aps,nofootinbib]{revtex4}
\usepackage[mathscr]{eucal}
\usepackage[dvips]{graphicx}
\usepackage{amsfonts,amssymb,amsthm}

\newcommand{\lalg}[1]{\mathfrak{#1}}  

\newcommand{\su}{\lalg{su}}

\newcommand{\so}{\lalg{so}}

\newcommand{\SU}{\mathrm{SU}}

\newcommand{\SO}{\mathrm{SO}}

\def \be{\begin{equation}}
\def \ee{\end{equation}}
\def \bes{\begin{eqnarray}}
\def \ees{\end{eqnarray}}

\newcommand{\Cb}{{\rm \bf C}}

\def \sl2{SL(2,\Cb)}
\begin{document}
\title{Group field theory and simplicial quantum gravity}
\author{\bf D. Oriti}
\affiliation{\small Institute for Theoretical Physics, Utrecht
University, Leuvenlaan 4, Utrecht 3584 TD, Netherlands, EU \\
and \\ Perimeter Institute for Theoretical Physics, 31 Caroline
St, Waterloo, Ontario N2L 2Y5, Canada \\ and \\ Albert Einstein
Institute, Am Muelenberg 4, Golm, Germany, EU \\ \\
doriti@perimeterinstitute.ca, daniele.oriti@aei.mpg.de}
\begin{abstract}
We present a new Group Field Theory for 4d quantum gravity. It
incorporates the constraints that give gravity from BF theory, and
has quantum amplitudes with the explicit form of simplicial path
integrals for 1st order gravity. The geometric interpretation of
the variables and of the contributions to the quantum amplitudes
is manifest. This allows a direct link with other simplicial
gravity approaches, like quantum Regge calculus, in the form of
the amplitudes of the model, and dynamical triangulations, which
we show to correspond to a simple restriction of the same.
\end{abstract}
\maketitle

\section{Introduction and motivation}
\subsection{GFTs, spin foams and simplicial gravity}
The field of non-perturbative quantum gravity is progressing fast
\cite{libro}, in several directions. Spin foam models \cite{SF}
are one of them, and can be understood as a covariant formulation
of the dynamics of loop quantum gravity \cite{LQG} and as a new
algebraic version of the discrete quantum gravity approach based
on path integrals, as for example Regge calculus \cite{williams}
and dynamical triangulations \cite{DT}. This line of research has
recently received further impetus with the introduction of new
models for 4-dimensional quantum gravity in
\cite{EPR1,EPR2,laurentkirillnew,eterasimone2}. The general idea
of spin foam models is to encode the kinematics of quantum gravity
in discrete quantum histories given by {\it spin foams}:
combinatorial 2-complexes labelled by group-theoretic data. The
2-complex is combinatorially dual to a simplicial complex, and the
algebraic data are interpreted as determining a possible
simplicial geometry, just as edge lengths do in traditional Regge
calculus. A quantum dynamics is specified by the assignment of a
probability amplitude to each spin foam, with the model being
defined by a sum over both 2-complexes and algebraic data
labelling them. At present the most complete definition of a spin
foam model is achieved by means of the so-called group field
theory formalism \cite{iogft,iogft2,laurentgft}. Group field
theories are quantum field theories over group manifolds (usually
the Lorentz group) characterized by a non-local pairing of field
arguments in the action, and can be seen as a generalization of
matrix models \cite{mm} that have proven so useful for our
understanding of 2d quantum gravity (and string theory) (and of
the subsequent, but less developed, tensor models
\cite{gross,ambjorn}). The combinatorics of the field arguments in
the interaction term of the GFT action matches the combinatorics
of (d-2) faces of a d-simplex, with the GFT field itself
interpreted as a (second) quantization of a (d-1)-simplex. The
kinetic term of the action, in turn, dictates the rules for gluing
two such d-simplices across a common (d-1)-simplex, and thus for
the propagation of (pre-)geometric degrees of freedom from one to
the next. See \cite{iogft,iogft2} for details. Because of this
combinatorial structure, the Feynman diagrams of a GFT are dual to
d-dimensional simplicial complexes, and are themselves given by
2-complexes. The field arguments assign to these 2-complexes the
same group-theoretic data that characterize spin foam models, and,
most importantly, the GFT perturbative expansion in Feynman
amplitudes define uniquely and completely \cite{mikecarlo} a spin
foam model. This fact alone makes GFTs a very useful tool, but
also leads to the suggestion that they can provide a more
fundamental definition of a dynamical theory of spin networks (as
a second quantized theory of spin networks, as clarified also in
\cite{ioetera}), in particular representing the best way to
investigate non-perturbative and collective properties of their
quantum dynamics \cite{gftfluid}. It can also be argued
\cite{iogft,iogft2,gftfluid} that GFTs represent a common
framework for both the loop quantum gravity/spin foam approach and
simplicial approaches, like quantum Regge calculus and (causal)
dynamical triangulations, whose basic ideas and structures they
incorporate.

\subsection{Motivation}
For this idea to be realized, or at least investigated in more
concrete terms, it is necessary to have at one's disposal a class
of group field theories whose Feynman (spin foam) amplitudes are
given by path integrals for simplicial gravity actions in 1st
order form (as appropriate for the type of variables appearing in
them). This has been partially achieved in \cite{iotim}. In this paper, we develop further, and in a sense bring to a completion,
the work in \cite{iotim} and present a group field theory for 4-dimensional quantum gravity, whose Feynman amplitudes are indeed
simplicial gravity path integrals with a clearly identified simplicial action
and with all the GFT variables possessing a clear geometric interpretation. More precisely, the model's amplitudes are path integrals for a discrete Plebanski action, i.e. a
discrete BF action with constraints on the $B$ variables that turn
them into a discretization of the continuum tetrad field
\cite{SF}, augmented by quantum corrections. On the one hand, therefore, we obtain an explicit link
with 4d simplicial quantum gravity, and with other simplicial
approaches. On the other hand, the ability to control the imposition of the constraints in a path integral context makes the proposed GFT a
nice step forward towards a spin foam formulation of 4d quantum
gravity, since the correct imposition of these constraints is the
declared goal of all current spin foam models
\cite{EPR1,EPR2,laurentkirillnew,eterasimone2} (we will discuss in the following both the good sides and the limitations of the model we are presenting). We achieve this by means of a generalization of the whole GFT
formalism, that allows to by-pass some not entirely satisfactory
(in our opinion) steps in the usual spin foam quantization
procedure, when re-phrased in a Lagrangian path integral language. Let us now sketch briefly these steps and the reasons
why we find them problematic.

\medskip

The classical inspiration is the Plebanski formulation
of gravity as a constrained BF theory \cite{SF}. 
This means that the theory one would really want to quantize in path integral
(Lagrangian) language is some discrete version of the Plebanski
action on a simplicial complex $\Delta$, schematically of the
form: \be Z_\Delta=\int\mathcal{D}g_{e}\mathcal{D}B_{f}\,\prod_{e}
C(B_{f\subset e})\,e^{i\sum_{f}tr(B_{f}F_{f}(g_{e\in\partial f}))}
\label{plebanski} \ee where we have localized the simplicity
constraints at the level of each tetrahedron (dual edge $e$) of
the simplicial complex \cite{laurentkirillnew} (see also
\cite{eteravalentin}). These constraints impose the restriction on
the {\it classical} discrete Lie algebra B variables giving
gravity from BF theory. Notice also that one should expect the presence, in the same path integral expression, of secondary second class constraints enforcing the consistency between the simplicity (Plebanski) constraints and the other constraints of the theory (following local Lorentz and diffeomorphism invariance, if not broken) \cite{sergei, sergeiothers}. These additional constraints are not clearly accounted for in recent spin foam models.

The usual spin foam procedure \cite{SF} mimics this path integral quantization but is applied at the level of quantum states directly, using methods from geometric quantization. The idea is the following. We do know the structure of states of quantum BF theory formulated as a spin foam model, the Ooguri-Crane-Yetter model, being the 4d counterpart of the Ponzano-Regge model for 3d gravity/BF), adapted to a simplicial complex: they are given by spin networks with links labeled by representations of $\SO(4)$ (in euclidean signature), with 4-valent vertices dual to tetrahedra of the corresponding simplicial complex (while their links are dual to the triangles of the same tetrahedra). We also know how to construct, starting from such states, an amplitude for a single 4-simplex that is then a function of the representations labeling its 10 triangles and the additional parameters labeling its 5 boundary tetrahedra. It is obtained by appropriate contraction of the 5 intertwining tensors associated to its 5 boundary tetrahedra. This is the main ingredient of the spin foam model defining the dynamics of the BF quantum states. The other contributions to the complete spin foam amplitude, associated to lower-dimensional simplices, i.e. triangles and tetrahedra, and usually thought of as \lq\lq measure terms\rq\rq can be similarly defined, but can be determined uniquely only by the requirement of triangulation independence (following from the topological character of quantum BF theory). The task for defining a spin foam dynamics for 4d gravity is then first of all to find appropriate restrictions on the $\SO(4)$ spin networks of BF theory, i.e. on their representation and intertwiner labels, that can be interpreted as the quantum analogue of the Plebanski constraints in the classical B variables of BF theory, and thus to characterize the spin network quantum states of 4d gravity. In the new spin foam models of \cite{EPR1,EPR2,laurentkirillnew,eterasimone2}, the crucial intermediate step is to re-write the same spin network states of BF theory in terms of coherent states of $\SO(4)$, whose defining parameters are interpreted as the quantum labels corresponding to the B variables. Second, one follows the steps leading to the BF spin foam amplitudes, but starting the newly defined candidate 4d gravity states, arriving at a proposal for the 4-simplex gravity amplitude. This procedure again leaves the measure terms in the spin foam model undetermined, and this time we do not expect, and thus we can not impose, any triangulation independence property.
Moreover, the identification of the analogue of the discrete classical B variables, e.g. with the parameters labeling the quantum coherent states, in the
quantized BF partition function or in its quantum states expressed in terms of group elements or group representations is not free of ambiguities\footnote{For example, as the parameters labeling the coherent states correspond to the mean value  only of the operators corresponding to the B variables, when computed in the same coherent states, one could argue that they can be identified with the same B variables only in some semi-classical approximation, and not in the fundamental theory; consequently, the Plebanski constraints should not be imposed on these parameters, as done in the new spin foam models, strongly.}. Consequently, the procedure for imposing the simplicity constraints on them, that would be rather straightforward at the path integral level, is itself ambiguous.

When re-phrased in terms of a path integral quantization, the same spin foam procedure
consists of four basic steps: 1) start from the BF path integral
discretized on a simplicial complex, with the $B$ field
discretized in terms of Lie algebra elements $B_t$ associated to
the triangles of the complex, and the connection degrees of
freedom encoded in group elements (to be thought of as parallel
transports of the same connection) associated to the edges of the
dual 2-complex and thus in 1-1 correspondence with the tetrahedra
of the triangulation; 2) integrate out the $B_t$ variables so that
one is left with a quantum amplitude function of group elements
only, and given by a delta function for each triangle (dual face)
imposing flatness of the corresponding holonomy (this is the
entire content of BF theory); 3) expand the delta function in
irreducible representations of the Lorentz group obtaining a
re-formulation of the original discrete path integral in terms of
group elements and representations of the same group; 4) identify
in this reformulation variables that can be argued to correspond
to the original Lie algebra variables $B_t$ and impose on them an
analogue of the Plebanski constraints on
the $B$ field, modifying the measure on such variables
in the partition function. This last step leaves us with a
candidate partition function for 4d quantum gravity, that can then
be put in a pure spin foam form by integrating out the group
variables. In the most recent spin foam models
\cite{EPR1,EPR2,laurentkirillnew,eterasimone2}, for example, a
basis of coherent states for the Lorentz group is used for
expressing the amplitudes in the representation picture, i.e.
after the expansion in step 3), their characterizing parameters
are identified with the discrete $B$
variables, and the simplicity (Plebanski) constraints are imposed
on them to give the final model. The initial BF path integral with
group elements and Lie algebra variables does not play an
essential role and one could have as well started directly from a
definition of BF theory as a product of delta functions on the
group, i.e. from the result of step 2)\footnote{The model presented in \cite{laurentkirillnew} was recently
given a simplicial path integral form in \cite{laurentflorian},
used in \cite{laurentflorian2} to study its semiclassical limit.
The resulting simplicial action is a non-standard BF-like action
involving explicitly discrete bivectors $B$ which are, however,
{\it defined} in term of the group representations and coherent
state parameters mentioned above, and not introduced as
independent Lie algebra elements and later identified with the
above in some approximation, as one would expect. So, the model
can still be understood in the context outlined above, and thus
subject to the same criticism. Concerning this last model as well as the problems in interpreting it in a path integral context, see the recent \cite{valentin2}}.

The corresponding group field theory construction of the same spin
foam models follows an analogous path (see \cite{SF} and, for the
new models, \cite{laurentkirillnew}). One
starts from a group field theory action designed to produce
Feynman amplitudes given by products of delta functions. The
action for such model (for $\phi\in\mathbb{R}$) is \cite{Ooguri}:
\bes S[\phi]\,&=&\,\frac{1}{2}\int dg_1 ...dg_4
\phi(g_1,g_2,g_3,g_4)\phi(g_4,g_3,g_2,g_1)\,+\,\nonumber \\ &-&\,\frac{\lambda}{5!} \int dg_1...dg_{10}
\,\left[\phi(g_1,g_2,g_3,g_4)\,\phi(g_4,g_5,g_6,g_7)\,\phi(g_7,g_3,g_8,g_9)\,\phi(g_9,g_6,g_2,g_{10})\,\phi(g_{10},g_8,g_5,g_1)\right]\;\;
. \ees The field is assumed to be symmetric under the diagonal
(right) action of the Lorentz group $G$ $\phi(g_i)=\phi(g_i
g)\;\;\forall g\in G$. This symmetry requirement is ultimately
responsible for the form of the quantum amplitudes. It introduces
a gauge connection on the dual links of the simplicial complex and
imposes flatness inside each 4-simplex, which, combined with the
trivial gluing condition imposed by the kinetic term, results in
imposing flatness on each dual face (triangle). Starting from the
above, the crucial step is the insertion of appropriate
constraints in the GFT action, in the above group picture (e.g.
for the Barrett-Crane model \cite{SF}) or, in the representation
picture, following step 3) and 4) above. For example, in the new
models \cite{laurentkirillnew} one can expand the GFT
kinetic term in group representations, and then in coherent states
of $G$, and finally constrain the corresponding parameters as the
simplicial geometry of a tetrahedron (corresponding to the GFT
field), when expressed in terms of bivectors $B_t$ suggests
\cite{eterasimone2}. Once more, there is no role for the Lie
algebra variables, that would directly correspond to the discrete
$B$ field.

\medskip

There are reasons to be dissatisfied with this procedure, and to look for alternatives, in particular to look for a proper Lagrangian formulation of a 4d spin foam model for gravity. 

We have already mentioned the difficulty in specifying uniquely the measure terms in the usual spin foam construction. A proper Lagrangian path integral derivation starting from BF theory (whose measure is unambiguous) could provide a prescription for these measure terms.

In a Lagrangian/path integral context, the two sets of Lie
algebra and group variables are classical and {\it independent} of
each other. Any relation between the two should arise in some
(dominant) configurations, satisfying the discrete equations of
motion. We want to reproduce exactly this feature, with a
corresponding \lq\lq doubling\rq\rq of independent geometric
variables, at the GFT level, with our generalised GFT formalism. Alternatively, the relation between the two should be in terms of some classical Fourier transform mapping one set of variables to the other. We will discuss briefly this possibility in the following.
On the other hand, the usual spin foam quantization introduces a quantum analogue of the B variables in terms of the Lie algebra generators acting on
function of the group variables, and thus lives from the start at the quantum (canonical) level, outside the path integral context.

Once an appropriate discretization of the simplicity constraints
has been chosen,  they are imposed
as delta function restriction in the path integral measure on the
classical independent B variables. This means that, contrary to steps 3) and 4) above, there is no ambiguity in how to impose the simplicity constraints, which is the point of debate in the usual spin foam construction. Furthermore, this also means that they have to be
imposed {\it before} performing the integration over these
variables, i.e. at the classical level, as they drastically affect
exactly this integration, contrary to the usual spin foam
procedure.

\medskip

In this work, we try to overcome the pitfalls of the
usual procedure when seen from a Lagrangian path integral point of view\footnote{Some of these motivations, and similar
technical issues, are shared by the work \cite{eteravalentin}
where a Lagrangian approach to a 4d gravity spin foam model on a
given simplicial complex is presented. Along the same line of reasoning, see the recent results of \cite{valentin, valentin2}, appeared after the completion of this work, and that we will discuss more in the following, as on the one hand they subtantiate further the above criticisms, and on the other hand clarify better some of our own results.}. We also make direct
contact with simplicial gravity at the level of the resulting
partition function (and not only as a guideline for the
construction). We work in (and develop further) a generalised
group field theory framework in which both group variables
(representing discrete connection variables) and Lie algebra
elements (representing discrete $B$ variables) are present. We
obtain an explicit simplicial gravity path integral for 4d gravity
in 1st order variables, with BF-like action and the simplicity
constraints manifestly implemented on the Lie algebra variables.
Moreover, thanks to the presence of both sets of variables, we can
keep under control and manifest the simplicial geometric picture
at all stages in the construction. This also allows us to
establish a direct link with other simplicial gravity approaches
like quantum Regge calculus, in the very form of the quantum
amplitudes of the model, and dynamical triangulations, which we
can show to correspond to a simple restriction of the same.

\medskip

The new model can be understood as a refinement of the class of
models introduced in \cite{iotim}. More details on the new 4d
gravity model, together with the construction and thorough
analysis of the simpler 3d gravity model can instead be found in
\cite{iotim2}. This work can be also understood in the context of
the line of research on the issue of causality in spin foam
quantum gravity and GFT, and on the construction of a unified GFT
framework for loop quantum gravity and simplicial quantum gravity,
developed in \cite{causal,feynman,generalised,iotim}.

\section{New GFT model for 4d quantum gravity}
\subsection{Definition of the model}

The action defining the classical dynamics of the
model is:
\begin{eqnarray*}
S &=&\frac{1}{2}\int \mathcal{D}x_i\mathcal{D}b^+_i \,\phi(x_i ; b^+_i) \, \mathcal{K}_m(x_i,b^+_i) \,\phi(x_i ; b^+_i) \,-\,   \frac{\lambda}{5!} \int \mathcal{D}x_{ij}\mathcal{D}b^+_{ij}\,
\, [P_g\phi](x_{1j} ; b^+_{1j}) .. [P_g\phi](x_{5 j} ; b^+_{5 j})\,
\mathcal{V}(x_{ij},b_{ij}) \,= \\ &=& \frac{1}{2}\int \mathcal{D}x_i\mathcal{D}b^+_i \,[P_BP_h\varphi](x_i ; b^+_i) \, \prod_i \left( \square_i^{S^3}\,+\, | b_i^+|^2 -\frac{m^2}{4}\right) \,[P_BP_h\varphi](x_i ; b^+_i) \,+\nonumber \\ &-&   \frac{\lambda}{5!} \int \mathcal{D}x_{ij}\mathcal{D}b^+_{ij}\,
\, [P_gP_BP_h\varphi](x_{1j} ; b^+_{1j}) .. [P_gP_BP_h\varphi](x_{5 j} ; b^+_{5 j})\,
\mathcal{V}(x_{ij},b_{ij}) \,= \\ &=& \frac{1}{2}\int \mathcal{D}g_i \mathcal{D}\tilde{g}_i\int\mathcal{D}b^+_i\mathcal{D}b_i^- \int\mathcal{D}\tilde{b}^+_i\mathcal{D}\tilde{b}_i^-\,\int\mathcal{D}h_i\mathcal{D}\tilde{h}_i\,\int\mathcal{D}N\int\mathcal{D}\tilde{N}\,\prod_i\delta\left( b_i^- + N b_i^+ N^{-1}\right) \,\delta(\sum_i b_i^+)\,\, \nonumber \\ &{}&\prod_i\delta\left( \tilde{b}_i^- + \tilde{N} \tilde{b}_i^+ \tilde{N}^{-1}\right) \,\delta(\sum_i \tilde{b}_i^+)\; \varphi(g_i h_i ; (b^+_i,b_i^-)) \,\left[ \prod_i \left( \square_i^{S^3}\,+\, | \tilde{b}_i^+|^2 -\frac{m^2}{4}\right)\delta(g_i\tilde{g}_i^{-1})\delta(b_i^+ - \tilde{b}_i^+) \right]\,\varphi(\tilde{g}_i \tilde{h}_i; (\tilde{b}^+_i,\tilde{b}_i^-)) \,+\nonumber \\ &-&   \frac{\lambda}{5!} \int \mathcal{D}x_{ij}\mathcal{D}b^+_{ij}\,
\, [P_gP_BP_h\varphi](x_{1j} ; b^+_{1j}) .. [P_gP_BP_h\varphi](x_{5 j} ; b^+_{5 j})\,=\nonumber \\ &=& \frac{1}{2}\int \mathcal{D}g_i \mathcal{D}\tilde{g}_i\int\mathcal{D}B_i \mathcal{D}\tilde{B}_i\,C(B_i)\,C(\tilde{B}_i)\varphi(g_i; B_i)\left[ \prod_i \int\mathcal{D}h_i\left( \square_i^{Spin(4)}+ \frac{1}{2}| \tilde{B}_i|^2 -\frac{m^2}{4}\right)\delta(g_ih_i\tilde{g}_i^{-1})\delta(b_i^+ - \tilde{b}_i^+) \right]\varphi(\tilde{g}_i ; \tilde{B}_i) +\nonumber \\ &-&   \frac{\lambda}{5!} \int \mathcal{D}x_{ij}\mathcal{D}b^+_{ij}\,
\, [P_gP_BP_h\varphi](x_{1j} ; b^+_{1j}) .. [P_gP_BP_h\varphi](x_{5 j} ; b^+_{5 j})\mathcal{V}(x_{ij},b_{ij}) \label{action}
\end{eqnarray*}

with the following ingredients:

\begin{itemize}

\item the fundamental field $\phi(x_1, b_1^+; x_2,b_2^+; x_3,b_3^+;x_4,b_4^+)$ lives on the cartesian product of
4 copies of
$S^3\times\su(2)$\footnote{We
deal here with the Riemannian version of the new model, and then discuss
briefly its (straightforwardly obtained) Lorentzian version.}. For our present concerns (but see the
discussion in \cite{iotim}), we can restrict our attention to real
fields. In turn, this field is obtained from a generic (real) field $\varphi(g_1,B_1;...;g_4,B_4)=\varphi(g_1,(b^+_1,b_1^-);...;g_4,(b^+_4,b_4^-))$ on $\left( Spin(4)\times\so(4)\right)^4$ by application of the two maps $P_B$ and  $P_h$.
In terms of these maps, as we are about to see in more details, one identifies the domain of dependence of $\phi$ $S^3$ as the quotient of the domain of $\varphi$ $Spin(4)$ by the diagonal $\SU(2)$ subgroup, and the $\su(2)$ algebra on which $\phi$ depends as the self-dual part of the $Spin(4)$ algebra on which $\varphi$ depends;

\item the measures of integration $\mathcal{D}g$, $\mathcal{D}h$ and $\mathcal{D}x$,  as the standard Haar measures on $Spin(4)$ and $\SU(2)$ respectively, and the corresponding induced measure on the homogeneous space $S^3\simeq Spin(4)/\SU(2)$, $\mathcal{D}B$  and $\mathcal{D}b$ are instead the standard
Lebesgue measure on the Lie algebra considered as a vector space.

\item the map $P_B$, acting on functions on $\so(4)^4$ as $P_B F(B_1,..,B_4) = P_B F( (b_1^+,b_1^-),...,(b_4^+,b_4^-)  ) = \prod_i\int\mathcal{D}b_i^- \int_{S^3}\mathcal{D}N\int\mathcal{D}\tilde{N}\,\prod_i\delta\left( b_i^- + N b_i^+ N^{-1}\right) \,\delta(\sum_i b_i^+) \,F( (b_1^+,b_1^-),...,(b_4^+,b_4^-)  ) = \prod_i\int\mathcal{D}b_i^- \mathcal{C}(B_i) F( B_1;....; B_4 ) $ constrains the four Lie algebra variables associated to the field
$\varphi$ imposing the simplicity constraints; its geometric meaning will be discussed below.

\item the projector $P_h$ maps $\varphi$ to a function on the
homogeneous space $\left(Spin(4)/\SU(2)\simeq S^3\right)^4$, and can be
obtained explicitly by group averaging
$P_h\varphi(g_1,B_1;...;g_4,B_4)=\int
dh_1..dh_4\,\varphi(g_1h_1,B_1;...;g_4h_4,B_4)$.

\item the kinetic operator is given by: \be
\mathcal{K}(g_i,B_i)\,=\,\prod_{i=1}^4 \Big (  \, B_i^2 \, + \,
\Box_{G_i} \,  - \frac{m^2}{4} \Big ) \label{kinetic}\ee with
$\Box_G$ the Laplacian (Dalambertian, in the Lorentzian case)
operator on the group manifold $Spin(4)$ or on the homogeneous space $S^3$, the square $|B|^2$ (or $|b_i^=|^2$) is taken using
the fundamental Killing form on the Lie algebra, and $m^2$ is an
arbitrary positive constant.

\item the projector $P_g$ imposes invariance under the diagonal
action of the group $Spin(4)$ on the group {\it and} Lie algebra
variables: $P_g\varphi(g_1,B_1;...;g_4,B_4)=\int dg\,\varphi(g_1g, g B_1
g^{-1};...;g_4 g,g B_4 g^{-1})$.

\item the vertex or interaction operator $\mathcal{V}$ is given by
\be \mathcal{V}(g_{ij},B_{ij})\,=\,\prod_{i \neq j = 1}^5
\delta(g_{ij} g_{ji}^{-1}) \delta(B_{ij} - B_{ji}).\ee

\end{itemize}

Let us now explain motivation and geometric content of all these
ingredients (for details, see \cite{iotim2}).

\medskip 

The key to the geometric interpretation of the various ingredients
and of the whole model is the geometric interpretation of the
field $\varphi(g_i,B_i)$ itself. We can think of it as the second
quantization of a tetrahedron whose geometry is characterized by
the four pairs of 1st order variables $g_i$ representing
elementary parallel transports of a Lorentz connection along paths
dual to the triangles of the tetrahedron, and $B_i$ representing
Lie algebra variables (or bivectors) associated to the same
triangles \cite{iogft,iogft2}. In addition, we assume that a
reference frame is associated to a tetrahedron and, thus, to each
field $\varphi$, and that the $B$
variables) are expressed in this frame.

\medskip

Let us now look at the geometric content of the interaction term.
First of all, as in other GFT models, the combinatorics of field
arguments is chosen to respect the combinatorics of triangles in a
4-simplex. It is this feature that characterizes GFTs as
peculiarly non-local field theories. Second, the symmetry imposed
by the projector $P_g$ amounts to the requirement that the
reference frame associated to each tetrahedron should be
arbitrary. In other words, one should be free to relate this frame
by means of an arbitrary parallel transport to some given, and
equally arbitrary, frame associated to the 4-simplex as a whole.
This parallel transport is effected by the group element $g$
integrated over in the projection $P_g$, that can be thought of as
associated to (half) dual edges in the 2-complex dual to the
triangulation to which the 4-simplex belongs. One can do the same
for all the tetrahedra/fields in the vertex term. This means that
their different variables all come from a single set of triangle
variables for the 4-simplex they belong to, then parallel
transported to different locations (the boundary tetrahedra) in
the simplex. This symmetry imposition amounts to a relaxation of
the usual symmetry imposed on the group variables alone in the
usual GFTs (and interpreted as imposing the closure constraint of
BF-like theories). The net result is that the field is not
invariant anymore under the diagonal action of the group, when
seen as a function of the group elements only, but it becomes {\it
covariant}, and this translates in representation space as a
covariance, as opposed to invariance, of the tensors between
the 4 representations associated to the triangles of the
tetrahedron. The need to relax this invariance to a covariance
requirement has been emphasized also, from a canonical
perspective, by Alexandrov \cite{sergei}. By a change of variables
one can move the group variables defining the projection operators
$P_g$ into the vertex function $\mathcal{V}$, then defined on
non-projected fields $\phi$'s (or $\varphi$). The result is a vertex function
given by: \bes \mathcal{V}(g_{ij},B_{ij})\,=\,\int \prod_{i=1}^5
dg_i \prod_{i \neq j = 1}^5 \delta(g_{ij}g_i g_j^{-1}g_{ji}^{-1})  \prod_{i \neq j = 1}^5 \delta(B_{ij} -
g_i^{-1}g_j B_{ji} (g_i^{-1}g_j)^{-1}).\ees The geometric meaning
is now transparent. The delta functions on the group impose that
the parallel transport $g_{ij}g_i g_j^{-1}g_{ji}^{-1}$ of the
Lorentz connection along the boundary of the wedge (portion of the
dual face inside each 4-simplex) associated to the triangle $ij$
shared by the tetrahedra labelled $i$ and $j$ and living inside
the 4-simplex that contains both these tetrahedra, is flat. This
is consistent with our piecewise-flat context. The deltas on the
Lie algebra impose that the Lie algebra variables (discrete B
field) associated to the same triangle in two different tetrahedra
are identified only after parallel transport from the center of
one tetrahedron (where they are originally defined) to the center
of the 4-simplex, i.e. to the single reference frame in which they
are in fact the same, and back to the other tetrahedron, with this
parallel transport effected by the group elements
$g_i$\footnote{This piecewise flat geometric requirements have
been also nicely discussed in \cite{EPR1,EPR2}, even though they
do not seem to be fully implemented in the quantization and thus
in the model presented there.}. We see therefore that the interaction term of the GFT model enforces a
symmetry requirements (local Lorentz invariance) and the trivial
kinematical geometry dictated by a piecewise-flat setting. 
Notice  that, because of the projectors $P_B$ and $P_h$ acting on the field $\varphi$, the variables $B_{ij}$ entering the vertex term will be constrained to be simple, and we will also have  $g_{ij}\in S^3$.

\medskip

Let us then give a look at the kinetic term. This is given by the
product of four Klein-Gordon operators acting on the group
manifold $G$ with a variable mass term given by $B^2$ plus an
arbitrary constant mass square shift, one for each triangle in the
tetrahedron the field refers to. This choice \cite{iotim}, first of all, relaxes the
identification between the discrete bivector $B$ associated to the triangle and
the Lie algebra generator $J$ seen as an operator acting on
functions of the group, and thus between its modulus square and
the (1st) Casimir of the algebra, which in turn is defined up to
an arbitrary constant shift. This specific choice is the simplest
Lorentz invariant one achieving the above. Obviously, others could
be considered. On the one hand, in the present context this step allows to deal with both $B$ and $g$ variables at the same time, it leads to a nice simplicial path integral, as we are going to see, and keeps the simplicial geometry of the model fully manifest. On the other hand, we may suspect already at this stage that it will lead us away from the dynamics of BF theory, even in absence of simplicity constraints, since the conjugate nature of these variables (characteristic of BF theory) is then not imposed in our model. 

The additional projectors $P_h$ lead to a
Klein-Gordon operator on the homogeneous space $S^3$. It results
in the amplitudes depending on a single angle parameter on the
group manifold. It is mainly motivated by the specific
discrete gravity action this leads to.
However, it also makes the same amplitudes, when expanded in
representations of the Lorentz group, involve only class 1
representations, as in the Barrett-Crane model and in the other
new spin foam models \cite{EPR1,EPR2,laurentkirillnew}. We know that this represents  partial imposition of the simplicity
constraints in terms of the connection variables, following their (partial) imposition on the Lie algebra generators identified with the discrete $B$ variables in a geometric quantization of BF theory . Here, on the other hand, the same constraints are
fully implemented on the Lie algebra variables $B$. Therefore, we could expect this additional restriction to be redundant, to some extent. We will discuss later on to what extent this is in fact true.

The simplicity constraints on the $B$
variables are imposed by the factor $\mathcal{C}(B_i)$
that is inserted for each field and thus for each
tetrahedron in the triangulation, {\it separately} in each 4-simplex to which the tetrahedron belongs. In fact, the GFT kinetic term
defines the gluing of two 4-simplices in the 4d triangulation
associated to each Feynman diagram (see \cite{iogft,iogft2})
across a common tetrahedron, represented by the two fields in the
GFT kinetic term (same tetrahedron seen in the two different
4-simplices). The simplicity constraints amount to the imposition
that the four bivectors associated to the four triangles $f$ of
the tetrahedron $t$ all belong to the same hypersurface (in flat
space), i.e. they are all normal to the same unit 4-vector,
interpreted as the normal to the tetrahedron \cite{sergei,
EPR1,EPR2, laurentkirillnew, eterasimone2}: $\exists N_t\in
S^3\,/\,\,B_f^{IJ} n_{t J} =0\,\,\forall f\subset t$. In this
case, the bivectors are interpreted as defining the so-called
\lq\lq area bivectors\rq\rq $A_f = e_1 \wedge e_2$ formed by the
wedge product of two edge (tetrad) vectors $e_{1,2}$ of the
triangle $f$. A dual version of the same constraints is the
requirement that the four bivectors are such that {\it their
duals} are normal to the same unit normal vector and thus belong
to the same hypersurface: $\exists N_t\in S^3\,/\,\,(*B_f)^{IJ}
N_{t J} =0\,\,\forall f\subset t$. In this case, they are
interpreted as {\it dual} to the same area bivectors: $B_f = *
A_f$\footnote{Note that, in absence of the Immirzi parameter, the two new spin foam models \cite{EPR1,EPR2,laurentkirillnew,eterasimone2}, based
on imposing the simplicity constraints on coherent states of the
Lorentz group, are distinguished exactly by the above
choice.}. To the above constraints, it is necessary to add the so-called closure constraint, imposing that the four bivectors associated to the triangles in a  given tetrahedron sum to zero, i.e. that the tetrahedron \lq\lq closes\rq\rq. Only when all these conditions are satisfied it is possible to invert the set of bivectors associated to the triangles of the simplicial complex for a set of tetrad vectors associated to the edges of the same and specifying a simplicial geometry. It is possible to show, moreover, that these set of constraints represents the discrete version of the continuum Plebanski constraints. The constraining factor $\mathcal{C}(B_i)$, brought in the action by the mentioned map $P_B$, has thus the
form: \bes \mathcal{C}(B_i)\,&=&\,\int_{S^3\simeq SU(2)} dN_t
\prod_{i=1}^{4}\,\delta^{(3)}( b_i^- \mp N_t \triangleright b_i^+
)\,\delta(\sum_i b_i^+)\,=\,\int_{S^3\simeq SU(2)} dN_t
\prod_{i=1}^{4}\,\delta^{(3)}( b_i^- \mp N_t  b_i^+ N_t^{-1})\,\delta(\sum_i b_i^+) = \nonumber \\ &=&\,\int_{S^3\simeq SU(2)} dN_t
\prod_{i=1}^{4}\,\int_{\mathfrak{su(2)}}db_i \,\,\delta( B_i -
(b_i, \pm N_t b_i N_t^{-1}))\delta(\sum_i b_i^+)\ees
where we have written the simplicity constraints using the
selfdual/anti-selfdual decomposition of the Lie algebra elements
$B_i$'s, and the identification of elements of the homogeneous
space $S^3$ with the $Spin(4)$ elements mapping the reference unit
vector $(1,0,0,0)$ into them, and further parametrized them as
$SU(2) \times SU(2)$ elements as $N_t = (N_t^+,N_t^-) = ( 1 ,
N_t)$ at the cost of a slight abuse of notation (see
\cite{laurentkirillnew}). The two possible signs $\mp$ correspond
to the first (second) way of imposing the simplicity constraints,
respectively. The arbitrariness of the normal vector $N_t$ is
enforced by the integration over $S^3$. In the last line, we have defined the simple
bivectors $b_i = (b_i, \pm b_i)$ in selfdual/anti-selfdual
decomposition, with similar abuse of notation, and with the same
sign ambiguity.

Notice that, due to the delta functions imposing the closure constraint, a (straightforward) regularization of the action and, later, of the amplitudes of the model will be needed.

Notice also that the anti-sefdual variables $b_i^-$ of the field $\varphi$ are not coupled among different fields, but are instead fixed completely as functions of the $b_i^+$ by the map $P_B$ (simplicity constraints) in each field separately. 

\subsection{Feynman amplitudes: discrete gravity path integrals}
The quantization of the model is obtained by the partition
function defined as a perturbative expansion in Feynman diagrams
in the coupling constant $\lambda$: \be Z = \int \mathcal{D} \phi
\,\, e^{i S[\phi]}\,\, =\,\, \sum_{\Gamma}
\frac{\lambda^{V_\Gamma}}{\textrm{sym}(\Gamma)} Z_{\Gamma}
\label{partition} \ee where $V_\Gamma$ is the number of vertices
in the Feynman diagram $\Gamma$ (see \cite{iogft,iogft2}),
sym($\Gamma$) is the order of automorphisms of the
diagram/complex, and $Z_{\Gamma}$ is the corresponding Feynman
amplitude, whose construction is detailed in \cite{iotim}, for
this general class of models, and in \cite{iotim2} for this
specific one and its 3d counterpart. The key ingredient is the
non-trivial propagator, the inverse of the kinetic term
\ref{kinetic}. Being this a product of Klein-Gordon operators on
the homogeneous space $S^3\simeq Spin(4)/SU(2)$ thanks to the
projection operators $P_h$, its inverse is taken to be the product
of Feynman propagator on the same homogeneous space, in turn equal
to the propagators on the group $SU(2)$, due to the isomorphism
between the two spaces, with variable mass given by $b_i^2
-\frac{m^2}{4}$. From now on we set $m^2=1$ because this
simplifies the resulting formulae. They can also be written in
terms of the analogous propagators on the full $Spin(4)$ group
acted upon by the same projectors $P_h$ (see \cite{camporesi}).
The resulting schematic form of the amplitudes is: \be
Z_\Gamma\,=\,\prod_{(ev)} \int_{Spin(4)} dg_{v e} \prod_{(ef )}
\int_\mathfrak{\su(2)} db^+_{ef }\int_{\su(2)}db^-_{ef}d\tilde{b}^-_{ef} \,\prod_{(ev)} \mathcal{C}(B_{f\subset
e; v})\,\prod_f A[ B_{ef} , g_{v e}] \label{amplitude}. \ee The
amplitudes factorize per dual face, apart from the measure factor
$\mathcal{C}$ imposing the simplicity constraints, which is associated to each dual edge $e$ (tetrahedron) for each vertex $v$ (4-simplex) it touches (belongs to). They depend on
one $\su(2)$ Lie algebra variable $b^+_{ef }$ for each tetrahedron $e$ sharing the
triangle dual to the face $f$, two similar (anti-selfdual) Lie algebra elements associated to the same tetrahedron-triangle pair, one for each 4-simplex $v$ sharing it, and one group element
$g_{ev}=(g_{ve})^{-1}$ for each half dual edge incident to a given
vertex $v$. Recall that all the anti-selfdual Lie algebra elements are fully fixed as functions of the self-dual ones, by the constraints $C$. Equivalently, one can think of a single $\so(4)$ Lie algebra element $B_{ef ; v} = (b_{ef}^+, b_{ef}^-)_v$ associated to each dual face (triangle) in each tetrahedron sharing it, within each one of the two 4-simplices $v$ to which the tetrahedron belongs, and with only the selfdual components of these Lie algebra elements being identified from one tetrahedron to the next, and across 4-simplices. The full holonomy around the dual face is given by $H
= \prod_{e\in\partial f} g_e$ (having chosen an arbitrary ordering
of the edges in the boundary of the face), where we have defined
$g_e= g_{ve}g_{ev'}$. They have the generic form: \be A[ B_{ef} ,
g_{ev}]\,= \, \mu(H_f , B_{ef}) \, \mathcal{W}(g_{ev}, B_{ef})\,
\, e^{i S^f_R [ B_{ef} , H_f ]} \, e^{i S_c^f [ B_{ef} , H_f ]} .
\label{faceamplitude}\ee We now discuss the various contributions
to these amplitudes.

\medskip

The first term to notice is the action term $S_R^f(B_{ef}, H_f) =
| B_{ef} | | [\theta_f ( H_f )]|$. $|B_{ef}|$ is the modulus of a
Lie algebra variable associated to the dual face $f$, and thus to
the corresponding triangle in $\Delta$, {\it in the reference
frame of one of the tetrahedra} sharing it. $\theta_f(H_f)$ is the
distance on the homogeneous space $S^3$ defined by the group
element $H_f$, which computes the simplicial curvature associated
to the dual triangle. The holonomy is defined starting from and
ending to the same tetrahedron in whose frame we defined $B_{ef}$;
the choice of the tetrahedron, and thus of the frame, is
immaterial \cite{EPR1,EPR2}. The notation $[..]$ indicates that
the amplitude depends only on {\it the equivalence class} of
distance angles corresponding to the same group element $H_f$;
this is a result of the periodic boundary conditions imposed on
the propagators on the compact manifold $S^3$. It is realized in
practice by defining this distance to be $\theta \pm 2\pi n$, with
$\theta \in [0, 2\pi]$ and $n\in\mathbb{N}$ identifying each of
the possible geodesics in $S^3$ on which the distance is computed,
and summing over $n$, thus adding the variable $n$ to the
configuration variables on which the amplitudes depend (see
\cite{iotim,camporesi} for details). Taking into account the
product over dual faces in (\ref{amplitude}), this gives an action
term associated to the whole triangulation $\Delta$ dual to
$\Gamma$ given by: $S_R = \sum_f S_R^f(B_f, H_f) = \sum_f | B_f |
| [\theta_f ( H_f )]|$. This action term characterizes the Feynman
amplitudes of the model as a simplicial gravity path integral. However, it is the presence of the terms
$\mathcal{C}_e$ and $\mathcal{W}_f$ that fully characterizes the
geometric content of this action and thus the dynamics of the
theory.

\medskip

The simplicity constraints $\mathcal{C}_{ev}(B_{f\subset e})$, as
discussed above, impose the geometric restriction that the set of
$B$ variables associated to the triangles of $\Delta$ are either
identified with or dual to the area bivectors for the same
triangles \cite{EPR1,EPR2,laurentkirillnew,eterasimone2}. These
constraints ensure that the set of $B_f$
variables can be put in bijective correspondence with a set of
tetrad vectors $E_l^I$ associated with the links of the
triangulation $\Delta$, and thus with a unique (up to local
Lorentz transformations) simplicial geometry
\cite{laurentflorian2}. In practice, they impose that each $B$
variable is of the form $(b_f, \pm N_e b_f N_e^{-1})$ for an
arbitrary vector $N_e$ common to all the bivectors associated to
the same tetrahedron $e$, and that tetrahedra close in the hypersurface orthogonal to $N_e$. It can be rephrased by saying that the
Lie algebra element $B_f$ is, up to the common Lorentz rotation
$N_e$, the generator of a $U(1)$ subgroup of the diagonal $SU(2)$
or of a corresponding anti-diagonal one, depending on the sign
chosen. The vector $N_e$ can be set to the form $(1,0,0,0)$ by
appropriate gauge choice in each tetrahedron (see
\cite{EPR1,EPR2,laurentkirillnew}).

\medskip

The contributions $\mathcal{W}(g_{ve}, B_{ef})$ are delta
functions on the Lie algebra. Choosing an ordering $1,...,N$ of
the $N$ edges in the boundary of the dual face, fixing an
arbitrary edge as the reference one and labelling it by $1$ (with
$N+1\equiv 1$), and defining $g_{e'e}=g_{e'v}g_{ve}$, they have
the form: \bes \;\;\;\mathcal{W}\,&=&\, \prod_{\bar{e}=1}^N
\delta(
B_{\bar{e}}\,-\,g_{1N}..g_{\bar{e}+1\bar{e}}..g_{21}\triangleright
B_1 )\,= \delta(B_N - g_{21}\triangleright
B_1)...\delta(B_1 - g_{1N}...g_{21}\triangleright B_1) \nonumber
\ees where we simplified the notation $B_{ef}\rightarrow B_e$ as
we are dealing with a single dual face.

These impose the obvious geometric requirement that the Lie
algebra variables appearing in the different tetrahedra but
corresponding to the same triangle, being the same fundamental Lie
algebra element just expressed in different reference frames, can
be obtained all from a single arbitrary one (here $B_1$) by
parallel transports of this across the various simplices sharing
the same triangle. Recall that the variables $B_{ef}$ are
integrated over in the discrete partition function. If this was a
free integration, then one could simply use the above deltas to
eliminate all the redundant $B$ variables, leaving in the end a
single one per dual face. This is what happens in the 3d case
\cite{iotim2}. However, the simplicity constraints $\mathcal{C}_{ev}$
impose restrictions on these integrals, and these restrictions,
together with the above deltas, imply further restrictions on the
parallel transport (connection) variables $g_{ev}$. We leave a
complete analysis of these restrictions to further work, but it is
clear that they impose a consistency or compatibility requirement
between simplicity constraints and gauge invariance, thus parallel
transport\footnote{Analogous compatibility conditions were noticed
in \cite{eteravalentin}.}.

Of particular
interest is the last of these deltas: \be \delta(B_1 -
g_{1N}..g_{21}\triangleright B_1)\,=\,\delta(B_1 -
H_f\triangleright B_1). \ee This imposes on the face holonomy the
condition $H_f\,= e^{i\theta_1 \hat{B}_e\,+\,i\theta_2
*\hat{B}_e}$, for arbitrary angles $\theta_{1,2}$; in other words,
it imposes that $H_f$ lives in the $U(1)\times U(1)$ Cartan
subgroup of $Spin(4)$ and, moreover, that this is aligned with the
subgroup generated by the two commuting Lie algebra elements
$\hat{B}_e=\frac{B_e}{|B_e|}$ and $*\hat{B}_e$.

\medskip

The geometric content of this condition is
revealed by taking the simplicity constraints into account as
well. First of all, notice that the presence of this delta
function reduces the gauge invariance of the Feynman amplitude
$Z_\Gamma$ from the invariance under $B_{ef}\rightarrow G B_{ef}
G^{-1}$, $H_f\rightarrow \bar{G} H_{f} \bar{G}^{-1}$ for arbitrary
$G,\bar{G}$\footnote{This follows from the fact that each
propagator contributing in building up the Feynman amplitude is
invariant under the conjugate action of the group $G$ on the
homogeneous space $S^3$ and that it depends on each $B$ only
through its modulus.}, that could be deduced from the action term
alone and the simplicity constraints (we will see that also the
other terms $\mu$ and $S_c$ would allow for this large symmetry)
to the smaller $B_{ef}\rightarrow G B_{ef} G^{-1}$,
$H_f\rightarrow G H_{f} G^{-1}$ for {\it the same group element
$G$}. This is indeed the symmetry of BF theory (see, for example,
\cite{EPR1,EPR2}). Consider now the simplicity constraints with
negative sign $B_{ef}=(b_{ef}, - n_e b_{ef} n_e^{-1})$ implying that
$B_{ef}$ is an area bivector $A_f$. This can be re-written as
$B_{ef} = N_e \triangleright * b_{ef}$ for the group element $N_e=
(1,n_e)$, with $n_e\in SU(2)$ \cite{laurentkirillnew}, and
$b_{ef}=(b_{ef},b_{ef})$ in selfdual/anti-selfdual notation. This
implies (simplifying again the notation) that $H_f\,= e^{i\theta_1
\hat{B}_f\,+\,i\theta_2 *\hat{B}_f}= N_e e^{i\theta_1
*\hat{b}_f\,+\,i\theta_2 \hat{b}_f} N_e^{-1} = N_e h_f N_e^{-1}$
with $h_f = e^{i\theta_1* \hat{b}_f\,+\,i\theta_2 \hat{b}_f}$. Let
us then take into account the symmetry noticed above. The
amplitudes, and in particular the action term depend on $H_f$ and
$B_f$ only up to their simultaneous rotation by an arbitrary group
element $G$. Let us choose this as $G=N_e$. Then it results that
the amplitudes are the same as those computed from $b_f$ and
$h_f$. More precisely, the action $S_R$ depends only on the
modulus of $b_f$, $|b_f|=|B_f|=|A_f|$ and on the distance on $S^3$
corresponding to the holonomy $h_f$. But $h_f=e^{i\theta_1
*\hat{b}_f\,+\,i\theta_2 \hat{b}_f}$, with $b_f = (b_f,b_f)=A_f$,
thus with $e^{i\theta_2 \hat{b}_f}$ belonging to the diagonal
$SU(2)$ subgroup of $Spin(4)$. Therefore we conclude that the
angle measuring the distance on $S^3$ and entering the action
$S_R$ is the component of $h_f$ along $*b_f=*A_f$, or of $H_f$
along $B_f$, i.e. $\theta_f(h_f)=\pm\theta_1$. We can therefore
interpret the contribution of each dual face to the simplicial
action as $|B_f||[\theta_f]|=tr(B_f F_f) = tr(B_f \ln{H_f})$
(the trace is in the Lie algebra) with the additional restriction
(that can be included in the measure of integration over the $B$
and $g$ variables by means of a simple Heaviside function)
$tr(B_f F_f)>0$. This same restriction has been argued for as a
(pre-)causality condition needed to define causal transition
amplitudes for BF theory and gravity in \cite{causalmatter3d}.
The
simplicity constraints imply that each $B$ variable comes from an
appropriate set of discrete tetrad variables. In turn, this
implies that the simplicial action $S_R$ can be understood
schematically as $S_R(E,g)=\sum_f tr(*A_f(E)F_f(g))$, i.e. as a
simplicial action for 4d gravity of a 1st order Regge type
\cite{magnea,laurentflorian,iotim}, in turn a discrete version of
the Palatini action for 4d gravity. Equivalently, this shows our
model to be a simplicial gravity path integral for a discrete
Plebanski formulation of 4d gravity as a constrained BF theory.

\medskip

A careful and complete analysis of the constraints $\mathcal{W}$, and of their interplay with the simplicity constraints $C$, has been performed in \cite{valentin}, after a first version of the present work had appeared. The analysis confirms the above and goes much further, in that it shows that these compatibility conditions  can be enough to solve completely the connection degrees of freedom in terms of the tetrad or bivector ones, and thus go from a 1st order formulation to a 2nd order one, which is fully equivalent to ordinary Regge calculus at the classical level. Moreover, the restriction on the connection degrees of freedom is such that the projection $P_h$ we have imposed on our fields, that leads to our quantum amplitudes to depend only on the $S^3$ component of the group elements, with the consequences we have discussed, become un-necessary. In fact, they would result in the component of the holonomy along $(b_f,b_f)$ to be trivialized altogether, obtaining the same form and geometric interpretation even when working with generic $Spin(4)$ group elements. However, at the same time the results of \cite{valentin} suggest that the correct way of imposing the simplicity constraints may differ slightly but crucially from the one we adopted here. Having imposed the simplicity constraints at the level of each field $\varphi$ independently, by means of the  map $P_B$, we end up with {\it two} normal vectors $N_{ev}\in S^3$ associated to the same tetrahedron $e$ one for each 4-simplex $v$, which our model treats as completely independent from each other. This has one important consequence. The restriction on the connection implied by the requirement that the two normals to two tetrahedra sharing the same triangle (either within the same 4-simplex or in two adjacent ones), both lie in the plane orthogonal to the area bivector associated to that triangle \cite{valentin}, is missing in our formulation. This requirement is important for the complete inversion of the connection variables as function of the bivector variables, thus for the definition of geometric dihedral angles from them, and it is not clear, at this stage, how much of this inversion one can still perform given our weaker form of the constraints. Notice also that this weaker form of the constraints is the same imposed in the Barrett-Crane model (within a  different formalism), and possible the origin of the problems faced by it (see also \cite{ioeteracoupling}). How the stronger form of constraints can be imposed in a GFT context is the subject of work in progress \cite{ioaristide2}.

\medskip

Finally, let us mention the last two contributions to the Feynman
amplitudes $Z_\Gamma$. The first is a measure term
$\mu(B_{ef},H_f)$ which is a real function of $|B_f|$ and $H_f$
only (with parametric dependence on the number of N dual
edges/vertices in the dual face $f$). The second is an additional
contribution $S_c$ to the classical action $S_R$ that we can
interpret as a quantum correction to the same. It has the same
dependence on the basic variables as $\mu(B,H)$. These two terms
are identified \cite{iotim} as the modulus and phase,
respectively, of the complex function:
\begin{eqnarray}
\;\;\;\;\;\;\;\;\nu( H, B, N) \propto  \frac{- i}{(N-1)!}
\frac{1}{\textrm{sin}(\theta(H_f))} \Bigg
( \frac{|[\theta(H_f)]|}{|B_f|}\Bigg )^{N-1}\, \sum_{K=0}^{N-2} \frac{(N + K
-2)!}{K! (N-K-2)!} \left(\frac{i}{ 2 |B_f| \, \, | [ \,
\theta(H_f) \, ] |}\right)^K\nonumber
\end{eqnarray}
Their explicit expressions and asymptotic form for
$|B_f|\rightarrow \infty$, $H_f\neq I$ are reported and discussed
in \cite{iotim,iotim2}. This analysis, which is the analogue of
the semi-classical expansion $j\rightarrow\infty$ performed in
usual spin foam models, is easily done {\it for arbitrary triangulation} thanks to the fact that the
full amplitude, modulo the various constraints (whose form is of
course unmodified in the asymptotic regime), i.e.
$\nu(B,H,N)e^{iS_R(B,H)}$, is given by an Hankel function, whose
asymptotic formulae can be then used. The end result is that
indeed the terms $S_c$ give subdominant large scale corrections
(of $1/R^n$ type) in this regime to the dominant Regge term.

\

In the end, the aim of obtaining a GFT model reproducing the
Plebanski formulation of gravity (\ref{plebanski}) in a simplicial
setting, and thus (implicitly\footnote{The re-writing of our
quantum amplitudes in pure spin foam form, i.e. as a function of
group representations only, can be straightforwardly obtained.
This has been done for the simpler but analogous models in
\cite{iotim}. While such re-writing can be useful, it does not
modify the content of the model.}) a new promising spin foam model
for 4d quantum gravity is realized.

\section{Discussion}
\subsection{Features of the model}
The Lorentzian version of the model can be constructed
straightforwardly. It amounts to: a) replacing the 4d rotation
group with $SL(2,\mathbb{C})$, and the $\mathfrak{g}$ with the
lorentz algebra; b) using, as target of the same projectors $P_h$
and as defining space for the normal vectors $N_t$ to the each
tetrahedron, the upper hyperboloid in Minkowski space
$\mathcal{H}^3\simeq SL(2,\mathbb{C})/SU(2)$. This corresponds to
considering only timelike normal vectors, i.e. tetrahedra embedded
in spacelike hypersurfaces only\footnote{The easiest guess on how to
generalise this Lorentzian model to one that includes both
spacelike and timelike tetrahedra would be to modify the
projectors $P_h$ to integrations over the $SL(2,\mathbb{R})$
subgroup of the Lorentz group, and the normal vectors $N$ to
vectors on the 3d de Sitter space $dS_3\simeq
SL(2\mathbb{C})/SL(2,\mathbb{R})$.}. The geometry of the vertex and
of the kinetic term of the GFT action is unchanged, as is the
geometric meaning of the Feynman amplitudes, reproducing once more
a Regge-Plebanski action.  Both the integration over a non-compact
space and the indefinite signature, however, give rise to
additional potential divergences, which have to be taken care of.

\medskip

In the 3d case, and in 4d in absence of the simplicity
constraints), the static-ultralocal truncation of the generalised
GFT models reproduces the usual 3d gravity spin foam model, i.e.
the Ponzano-Regge model, and the analogous spin foam model for 4d
BF theory \cite{iotim,iotim2}. This amounts to maintaining the
same field varibles, symmetries and interaction, but dropping the
derivative (and variable mass) terms in the kinetic term, being left with simple delta functions. One can consider here a similar truncation, and analyse the corresponding Feynman amplitudes. 

Because of the projections $P_h$, and because of the constraints on the connection following the imposition of simplicity on the bivectors $B$, that we discuss below, the amplitudes will not force
the dual face holonomies to be flat, as in BF. The simplicity
constraints $\mathcal{C}(B)$ will still restrict the Lie algebra
elements to be simple, and, in conjunction with the symmetry
projectors $P_g$ will still imply further restrictions
$\mathcal{W}$ on the connection variables ensuring compatibility
between simplicity conditions and parallel transport. Because of
the projections $P_h$ the resulting model can be expected to be similar to the
Barrett-Crane spin foam model, also due to the weak form of the simplicity constraints we have chosen. However, because of the mentioned
compatibility restrictions, thus, because of the modified symmetry
requirement $P_g$, we may still obtain a
different model. In particular, we expect a stronger correlations
among 4-simplices, effected by the matching of $B$ variables (and
not only their modulus) across them, even if the connection variables would turn out to be not sufficiently constrained to be geometric. 

The compatibility conditions $\mathcal{W}(B,g)$ are another aspect
of the new model that need further study. First of all, one needs
to clarify which restrictions they impose, exactly, on connection
degrees of freedom, as we have discussed. 
Most important, maybe, is a clearer understanding of their
geometric meaning and of their interpretation from the point of
view of the canonical quantization of Plebanski action. One could
in fact conjecture a relation between these constraints and the
secondary second class constraints identified in
\cite{sergeiothers} and whose importance for the spin foam
quantization has been emphasized in \cite{sergei}. In particular,
in \cite{sergei} it is argued that the presence of these secondary
second class constraints is a necessary consequence of the
simplicity conditions when considered together with the relaxed
Lorentz covariance of the intertwiners. Moreover, \cite{sergei}
argues that these constraints would be essential for a proper
integration over the connection variables by modifying the path
integral measure with respect to the one in BF theory. Both these
origin, features and consequences are shared by our constraints
$\mathcal{W}(B,g)$, thus supporting our conjecture. In fact, they are (at least in the stronger form of \cite{valentin}) equivalent in terms of resulting restrictions on the connection to the discrete gluing conditions (\lq\lq edge simplicity\rq\rq) identified and studied in \cite{biancajimmy}. 

The quantum corrections $S_c$ to the classical Regge action $S_R$
deserve more study as well. Three features can be already noticed.
First of all, they follow directly from the choice of GFT action,
that therefore somehow fixes not only which terms will appear in
the large distance limit or in any other approximate regime of the
amplitudes, but also their relative coefficients. This is also
true for the corrections appearing in the $j\rightarrow\infty$,
$H_f\approx Id$ limit of the simplicial action defined from
coherent states in \cite{laurentflorian2}. Second, this
approximation would be the analogue of the approximation
$|B_f|\rightarrow\infty$, $H_f\approx Id$ of our full action $S_R
+S_c$. In order to analyze this regime, however, a more careful
study than the one in \cite{iotim}, that effectively applies to
the regime $|B_f|\theta_f(H_f)>>1$ only, is needed. Third, it can
be shown \cite{iotim2} that these corrections originate directly
from the gluing of 4-simplices, i.e. from the off-shell
propagation of $B$ and $g$ variables across them. In fact, the
amplitude for a single 4-simplex with boundary has a similar
simplicial path integral form, depends exactly on the variables
characterizing BF/Plebanski theory on a manifold with boundary,
with fixed-B boundary conditions, but involves the action $S_R$
only \cite{iotim2}. More generally, it is not clear, at this stage
of development, whether the presence of the quantum corrections
$S_c$, and in general the precise form of the measure $\nu$ is to
be considered correct from the geometric point of view. It is
possible that the GFT model itself will have to be
amended in order to eliminate the corrections $S_c$ altogether and
thus simplify the measure $\nu$.

An interesting {\it extension} of the model would be
instead one including the Immirzi parameter in the quantum
amplitudes, and thus giving a quantization of (the simplicial
version of) the Holst action. This parameter, crucial for the LQG
formalism, enters prominently in the new spin foam models
\cite{EPRL,laurentkirillnew}, and affects significantly both their
kinematics and dynamics. One would expect the Immirzi parameter to be straightforwardly introduced by modifying the simplicity map $P_B$ only. This generalized map involving the Immirzi parameter is easy to introduce, indeed, but this modification does not seem to be enough to achieve the expected form of the action in our simplicial path integral. The reason seems to be that a modification of the map $P_h$ seems needed as well, and in turn this can be understood as due to our weaker imposition of the simplicity constraints. We stress again that, indeed, with a stronger imposition \cite{valentin} of the same, the maps $P_h$ could be avoided altogether and the connection would be completely fixed as a function of the $B$'s. What is less clear at this stage is what should
substitute the projections $P_h$ or, in other words, what
restrictions on the connection degrees of freedom does the
inclusion of the Immirzi parameter imply, given our weaker constraints. This is actually rather
unclear also in the usual spin foam models
\cite{EPRL,laurentkirillnew,ioeteraImm}.

\medskip

While all of the above concerns the analysis or improvement of our model within the same generalized GFT formalism, an altogether different line of development is suggested by several of its features. 

This new direction \cite{ioaristide1,ioaristide2} stems from the idea that, instead of relaxing the conjugate nature (at the classical level) of $B$ and $g$ variables in the definition of the model, treating them on equal footing as arguments of the GFT field, as we do here, one could instead take this conjugacy relation as the starting point for introducing the $B$ variables in the GFT formalism. This would mean to {\it map} the usual group field theories in which the field is a function of group elements into {\it non-commutative and non-local field theories} on several copies of the corresponding Lie algebra, using a generalized (non-commutative) Fourier transform, of the type developed in \cite{laurent-majid}. A first motivation for doing so is that, in the model we have presented, we are dealing with the Lie algebra of $\so(4)$ (and its subalgebras) as  an ordinary vector space, thus neglecting its non-commutative nature; this would lead to suspect that, in doing so, we are neglecting some relevant information that should instead be incorporated i a correct model using the $B$ variables as arguments of the GFT field. More precisely, The non-commutative nature of the algebra would result in a non trivial star product for functions on it, when seen as functions on $\mathbb{R}^3$ (for $\su(2)$) or $\mathbb{R}^6$ (for $\so(4)$), as we do here. Moreover, the use of a non-trivial star product for the products of fields in the GFT action would result in a different composition rule for vertex amplitudes with propagators in the construction of the Feynman amplitudes of the model, and we have already noticed above how the rather puzzling quantum corrections $S_c$ to the BF action appearing in the model we have presented are the result of the way our individual vertex (4-simplex) amplitudes (itself given by a simple BF action)  compose. Also, notice that the maps $P_B$ could be turned into true projections, in a non-commutative setting, and one could avoid any divergence resulting from them, in particular from the use of an ordinary delta function to impose the closure constraint; this would be due to the fact that the non-commutative delta function on the Lie algebra \cite{laurent-majid} is a perfectly regular function when seen as a function on $\mathbb{R}^3$ (for $\su(2)$) or $\mathbb{R}^6$ (for $\so(4)$). The step to a non-commutative setting could then be motivated from the point of view of regularization only, even in absence of the other mentioned motivations. Finally, we mention another hint for the need to move to a non-commutative setting, arising from the analysis of the model we have presented. It has recently been shown, both in 3d and in 4d \cite{emergent}, that effective scalar field theories on flat non-commutative spaces emerge naturally from group field theories, with the group manifold underlying the GFT providing the momentum degrees of freedom of the effective matter field, while the non-commutative position variables are expected to be related to the conjugate $B$ variables. A similar analysis performed using the generalized GFT formalism we have presented in this paper shows \cite{ioalessandro} that, while everything works fine even for these generalised GFT models as long as one focuses on the group sector, the expected matter field theory fails to emerge if one focuses instead on the Lie algebra sector of the same GFT models. This failure can be indeed traced back to having neglected the non-commutative structure of the same Lie algebra at the level of the fundamental GFT, i.e. in the definition of the generalised formalism itself.

At the same time, we think we should expect several features of the model we have presented to show up also in any new model using the $B$ variables and constructed by means of non-commutative tools. For example, the way the simplicial geometry is implemented at the level of the action and in particular the way the simplicity constraints are implemented, the interplay between simplicity constraints and gauge invariance, and the consequent compatibility relations encoded in the contributions $\mathcal{W}$  to the feynman amplitudes.

\subsection{The dynamical triangulations sector}
Finally, let us conclude by showing a direct link between our new
GFT model and the dynamical triangulations approach. This is made
possible by the manifest simplicial geometric meaning of variables
and amplitudes, that we can now put to use. We do not consider any
additional causality restriction that one could impose on the
model in order to obtain a restricted class of Feynman
diagrams/triangulations summed over. These restrictions are
crucial for the dynamical triangulations approach in its most
modern form, called indeed {\it causal dynamical triangulations}
\cite{DT}. Unfortunately, a field theoretic understanding of them
in 4d in terms of some generalization of matrix models, and thus
at the GFT level, is lacking, having been only recently achieved
in the 2d case \cite{dariojoe}. Consequently, we can make a link
only with the \lq\lq old\rq\rq dynamical triangulations approach:
we obtain a model defined as a sum over equilateral triangulations
(of arbitrary topology) weighted by a simplicial gravity action
(Regge action plus corrections). The 4d DT approach is based on a
{\it 2nd order} formulation of gravity where both area of
triangles and dihedral angles between (d-1)-simplices are
functions of {\it fixed} edge lengths. Our GFT model
(\ref{action}) corresponds to a {\it 1st order} formulation where
areas are given by $|B_f|$ and dihedral angles are encoded in the
connection variables $g_e$. Therefore, in order to reproduce a DT
setting we have to: a) kill all connection degrees of freedom, in
turn fixing the local Lorentz invariance, and do so in such a way
as to obtain dihedral angles in each 4-simplex corresponding to
the values $\phi$ obtained from the edge lengths in the
equilateral case; b) kill \lq\lq direction\rq\rq degrees of
freedom in the B's and fix their modulus to a constant $|B_f|$
equal to the value of the area of triangle $f$ again in the
equilateral case. Note that clearly this means having taken
already into account, in both the $B$ and $g$ variables, the
simplicity constraints. A candidate model for a GFT formulation of
dynamical triangulations\footnote{It is also possible that
restriction to the dynamical triangulations sector of our full GFT
model is achieved {\it dynamically}, without an explicit
restriction of the initial model \cite{iotim}} (in the Riemannian
case\footnote{The model can be extended to the Lorentzian case by
distinguishing timelike and spacelike dihedral angles and triangle
areas.}), or for the DT subsector of our GFT model, is: \[ S = \frac{1}{2} \int \mathcal{D}g\,\phi(g_i)
\mathcal{K}_{|B|}(g) \phi(g_i) \,+\, \frac{\lambda}{5!} \int
\mathcal{D}g\,\phi(g_{1j}) ... \phi(g_{5 j})\, \mathcal{V}_h(g) \]
for $\mathcal{K}_{|B|}(g_i)=\prod_{i=1}^4 \Big ( |B|^2  +
\Box_{G_i} - \frac{1}{4} \Big ) $, with fixed $|B|$ and
$\mathcal{V}_h(g)\,=\,\prod_{i \neq j = 1}^5 \delta(g_{ij} \,h\,
g_{ji}^{-1})$ with fixed $h = e^{i\phi} \in U(1)$. From this
action, as it can easily be verified, one derives Feynman
amplitudes of the same type of (\ref{amplitude}) without the
constraining factors $\mathcal{C}$ and $\mathcal{W}$ nor any
integration over the group or Lie algebra variables. The area of
every triangle will be $|B|$ and the angle of holonomy will be
$\theta_f = N_f \phi$, with $N_f$ the number of 4-simplices
sharing the triangle $f$. These parameters $|B|$ and $\phi$ can be
of course chosen to match those computed from the edge lengths of
an equilateral triangulation. The above model still defines a
group field theory, but the group variables $g_i$ have the only
effect of ensuring the gluing of simplices in the construction of
the Feynman diagrams/amplitudes, and then disappear from the same
(in the bulk, while they will still label boundary states whose
kinematical variables, however, will not have any dynamics). One
obtains therefore a sum over equilateral triangulations weighted
by an amplitude involving the Regge action and depending on the
above two parameters. When taken together with the weighting
factor depending on the coupling constant $\lambda$ in
(\ref{partition}) one notices that {\it the GFT coupling constant
can be interpreted as the exponential of ($i$ times) the
cosmological constant}, that ends up multiplying the number of
4-simplices in the triangulation, matching exactly the usual DT
construction. A Wick rotation can then easily devised, as in the
DT approach to give a euclideanized partition function, suitable
for numerical analysis. The main differences from the usual
dynamical triangulations approach are: a) the Regge term is
augmented by the quantum corrections term $S_c$, function of the
same parameters, also resulting from b) the measure term $\nu$
depending on the number of simplices per triangle. These terms
have to be thoroughly analyzed, but it is a rather general result
in the DT approach that its main features do not depend heavily on
the details of the measure or of the action chosen, being rather
dictated by entropic factors \cite{DT}. c) The holonomy angle
$\theta$ still enters the amplitudes through its equivalence class
$[\theta_f]$ (in fact, the expression for the deficit angle in
terms of the holonomy angle obtained from dihedral angles would be
$\epsilon_f=2\pi - \theta_f$). This is a direct consequence of our
group-theoretic framework, with fields and propagators defined on
a compact group manifold (notice that this feature is not present
in the Lorentzian setting for spacelike triangles, whose
corresponding holonomy is given by a boost parameter). This model
will be analyzed elsewhere.

However, it already allows us to consider two scenarios,
concerning the relation between GFTs and the dynamical
triangulations approach.

I) The reduced GFT model matching the dynamical triangulations
amplitudes may be still too general to achieve a good continuum
limit. This would be the perspective coming from the dynamical
triangulations side. In fact, first of all, it lacks the causality
restrictions we mentioned above and that proved so crucial in
recent developments: it still sums over different topologies and
it does not incorporate any foliation structure for the
triangulations of trivial topology. Second, it does include, in
the configurations summed over, pseudo-manifolds, i.e. manifolds
with conical singularities, whose characterizing features are also
not understood at the field theoretic level. In this situation,
with trivial amplitudes associated to the simplicial complexes,
entropic considerations dominate and the model is likely too
pathological.

However, a second perspective is possible, coming from our GFT
approach. II) One could argue that the model is already too
restricted, and that its pathologies are possibly cured within the
full model instead. In particular, the presence of non-trivial
amplitudes is crucial for both symmetry and renormalization
considerations. The restricted model implies a trivialization of
the gauge symmetries of the full one, with a consequent
trivialization of the Ward identities, that lead to non-trivial
relations among amplitudes associated to different Feynman
diagrams (simplicial complexes). Complexes that would not be
counted twice in the full sum, because of symmetries, are now
distinguished in terms of their amplitudes depending on pure
combinatorics. Moreover, the non-trivial amplitudes can be crucial
in implementing a good renormalization procedure, and in devising
suitable approximate regimes in which manifold-like configurations
dominate with respect to pseudo-manifold ones. Recent results
in 3d confirm this possibility \cite{renorm}.

\section{Conclusions}
We have presented a new GFT/spin foam model for 4d quantum
gravity. It is based on a recent extension of the GFT formalism,
in which the field depends on both group and Lie algebra
variables, representing the discrete analogue of the variables in
a BF-like formulation of gravity. The new model allows a
straightforward implementation of the simplicity constraints that
give gravity from BF theory, and has quantum amplitudes with the
explicit form of simplicial path integrals for gravity.
In doing so it sidesteps some ambiguous aspects of the usual spin
foam quantization procedure. The geometric interpretation of the
variables and of the different contributions to the quantum
amplitudes is also made manifest. Moreover, thanks to the links
with other discrete approaches to quantum gravity, that we have
explicitly shown, its usefulness and interest may well extend
beyond the spin foam or loop quantum gravity framework. While
there are still several aspects of the new model that should be
analyzed further, and interesting variations of the construction
leading to it that can be considered, we believe that the new
model represents an altogether new approach to the spin foam
quantization of gravity and opens up a whole landscape possible
new developments, some of which we have discussed in some detail.

\section*{Acknowledgements}
We thank T. Tlas, with whom part of this line of research has been
developed, for many useful discussions, criticisms and
suggestions. Special thanks are due to V. Bonzom and M. Smerlak, and to S. Alexandrov, for their interest, insightful criticisms and help. We also thank A. Baratin, B. Dittrich, J. Engle,
E. Livine, K. Noui and C. Rovelli for comments on
this work. Financial support from the A. Von Humboldt Foundation
through a Sofja Kovalevskaja Prize is gratefully acknowledged.


\begin{thebibliography}{99}
\bibitem{libro} D. Oriti, ed., {\it Approaches to Quantum Gravity}, Cambridge
University Press, Cambridge (2009)
\bibitem{iogft} D. Oriti, in \cite{libro}, [arXiv:
gr-qc/0607032]
\bibitem{iogft2} D. Oriti, in {\it Quantum Gravity}, B. Fauser, J. Tolksdorf and E. Zeidler, eds.,
Birkhaeuser, Basel, (2007), [arXiv: gr-qc/0512103]
\bibitem{laurentgft} L. Freidel, Int.J.Phys. \textbf{44}, 1769-1783, (2005) [arXiv: hep-th/0505016]
\bibitem{mm} F. David,  Nucl. Phys. B257, \textbf{45} (1985); P. Ginsparg, \lq Matrix models of 2-d gravity', [arXiv: hep-th/9112013]
\bibitem{gross} M. Gross,  Nucl. Phys. Proc. Suppl. \textbf{25A}, 144-149, (1992)
\bibitem{ambjorn} J. Ambjorn, B. Durhuus, T. Jonsson,  Mod. Phys. Lett. \textbf{A6}, 1133-1146, (1991)
\bibitem{boulatov} D. Boulatov,  Mod. Phys. Lett. \textbf{A7}, 1629-1646, (1992), [arXiv:  hep-th/9202074]
\bibitem{mikecarlo} M. Reisenberger, C. Rovelli,  Class. Quant. Grav. \textbf{18}, 121-140, (2001), [arXiv: gr-qc/0002095]
\bibitem{gftfluid} D. Oriti, Proceedings of Science, [arXiv:0710.3276]
\bibitem{SF} D. Oriti,  Rept. Prog. Phys. \textbf{64}, 1489, (2001), [arXiv: gr-qc/0106091]; A. Perez, \lq Spin foam models for quantum gravity', Class. Quant. Grav. \textbf{20}, R43, (2003), [arXiv: gr-qc/0301113]
\bibitem{LQG} C. Rovelli, {\it Quantum Gravity}, Cambridge University Press, Cambridge, (2006)
\bibitem{ioetera} E. Livine, D. Oriti, {\it Second quantization of spin network wave functions and quantum flat 3d space}, to appear
\bibitem{williams} R. Williams, in \cite{libro}
\bibitem{DT} J. Ambjorn, J. Jurkiewicz, R. Loll,  Phys.Rev.D \textbf{72}, 064014, (2005), [arXiv: hep-th/0505154] ; J. Ambjorn, J. Jurkiewicz, R. Loll, Contemp. Phys. \textbf{47}, 103-117, (2006), [arXiv: hep-th/0509010]
\bibitem{Ooguri} H. Ooguri, Mod. Phys. Lett. A \textbf{7}, 2799-2810 (1992), [arXiv: hep-th/9205090]
\bibitem{iotim} D. Oriti, T. Tlas, Class. Quant. Grav. 25, 085011 (2008), [arXiv:0710.2679]
\bibitem{causal} E. Livine, D. Oriti,  Nucl. Phys. B \textbf{663}, 231 (2003), [arXiv: gr-qc/0210064]
\bibitem{feynman} D. Oriti,  Phys. Rev. Lett. \textbf{94}, 111301 (2005), [arXiv: gr-qc/0410134]
\bibitem{causalmatter3d} D. Oriti, T. Tlas, Phys.Rev. D \textbf{74}, 104021, (2006), [arXiv: gr-qc/0608116]
\bibitem{generalised} D. Oriti,  Phys.Rev. D\textbf{73}, 061502 (2006) [arXiv: gr-qc/0512069]
\bibitem{iotim2} D. Oriti, T. Tlas, {\it Encoding simplicial geometry in group field theories}, ITP-UU-08/60, SPIN-08/47, to appear;
\bibitem{laurentkirillnew} L. Freidel, K. Krasnov, Class. Quant. Grav. \textbf{25}, 125018 (2008) [arXiv: 0708.1595]
\bibitem{magnea} M. Caselle, A. D'Adda, L. Magnea,  Phys.Lett.B \textbf{232}, 4, 457 (1989); J. W. Barrett, Class.Quant.Grav. \textbf{11}, 2723-2730 (1994), [arXiv: hep-th/9404124]
\bibitem{EPR1} J. Engle, R. Pereira, C. Rovelli, Phys. Rev. Lett. \textbf{99}, 161301 (2007), [arXiv:0705.2388]
\bibitem{EPR2} J. Engle, R. Pereira, C. Rovelli, Nucl. Phys. B \textbf{798}, 251 (2008), [arXiv: 0708.1236]
\bibitem{eterasimone2} E. Livine, S. Speziale, Europhys. Lett. \textbf{81}, 50004 (2008),  [arXiv: 0708.1915]
\bibitem{laurentflorian} F. Conrady, L. Freidel, Class. Quant. Grav. \textbf{25}, 245010 (2008), [arXiv:0806.4640]
\bibitem{laurentflorian2} F. Conrady, L. Freidel, [arXiv:0809.2280]
\bibitem{eteravalentin} V. Bonzom, E. Livine, [arXiv:0812.3456]
\bibitem{sergei} S. Alexandrov, Phys. Rev. D \textbf{78}, 044033 (2008), [arXiv:0802.3389]
\bibitem{camporesi} R. Camporesi, Phys. Rept. \textbf{196}, 1
(1990)
\bibitem{sergeiothers} S. Alexandrov, E. Buffenoir, P. Roche, Class. Quant. Grav. \textbf{24}, 2809 (2007), [arXiv: gr-qc/0612071]; E. Buffenoir, M. Henneaux, K. Noui, P.
Roche, Class. Quant. Grav. \textbf{21}, 5203 (2004), [arXiv:
gr-qc/0404041]
\bibitem{EPRL} J. Engle, E. Livine, R. Pereira, C. Rovelli, Nucl. Phys. B \textbf{799}, 136 (2008), [arXiv:0711.0146]
\bibitem{ioeteraImm} E. Livine, D. Oriti, Phys. Rev. D \textbf{65}, 044025
(2002), [arXiv: gr-qc/0104043]
\bibitem{ioeteracoupling} E. Livine, D. Oriti, JHEP, 0702:092 (2007) [arXiv: gr-qc/0512002]
\bibitem{valentin} V. Bonzom, Class. Quant. Grav. \textbf{26}, 155020 (2009) [arXiv:0903.0267]
\bibitem{valentin2} V. Bonzom, [arXiv:0905.1501]
\bibitem{biancajimmy} B. Dittrich, J. Ryan, [arXiv:0807.2806]
\bibitem{ioaristide1} A. Baratin, D. Oriti, {\it The non-commutative representation for group field theories and simplicial gravity}, to appear
\bibitem{ioaristide2} A. Baratin, D. Oriti, {\it A new group field theory for 4d quantum gravity}, to appear
\bibitem{laurent-majid} L. Freidel, S. Majid,  Class. Quant. Grav. \textbf{25}, 045006 (2008) [arXiv: hep-th/0601004]
\bibitem{emergent} D. Oriti, J. Phys. Conf. Ser. \textbf{174}, 012047 (2009) [arXiv:0903.3970]
\bibitem{ioalessandro} A Di Mare, D. Oriti, {\it Emergent non-commutative matter from a quantum geometric background}, in preparation
\bibitem{dariojoe} D. Benedetti, J. Henson, [arXiv:0812.4261]
\bibitem{renorm} L. Freidel, R. Gurau, D. Oriti, Phys. Rev. D \textbf{80}, 044007 (2009), [arXiv:0905.3772]
\end{thebibliography}
\end{document}